\documentclass[submission,copyright,creativecommons]{eptcs}
\providecommand{\event}{TARK 2021} 
\usepackage{breakurl}             
\usepackage{underscore}           

\usepackage{graphicx}
 \usepackage{placeins}

\usepackage{tikzfig}

\usepackage{stmaryrd}

\usepackage{extarrows}
\usepackage{slashbox}

\usepackage{tkz-graph}
\GraphInit[vstyle = Shade]
\tikzset{
  LabelStyle/.style = { rectangle, rounded corners, draw,
                        minimum width = 2em, fill = yellow!50,
                        text = red, font = \bfseries },
  VertexStyle/.append style = { inner sep=5pt,
                                font = \Large\bfseries},
  EdgeStyle/.append style = {->, bend left} }

\usepackage{lineno,hyperref}
\usepackage{algorithm}
\usepackage[noend]{algorithmic}
\usepackage{amssymb}
\usepackage{graphicx}
\usepackage{amsmath}
\usepackage{algorithm}
\usepackage[noend]{algorithmic}
\usepackage{url}
\usepackage{caption}
\usepackage{multirow}
\usepackage{threeparttable}
\usepackage{color}
\usepackage{lineno}

\usepackage{tkz-graph}

\tikzset{
  LabelStyle/.style = { rectangle, rounded corners, draw,
                        minimum width = 2em, fill = yellow!50,
                        text = red, font = \bfseries },
  VertexStyle/.append style = { inner sep=5pt,
                                font = \Large\bfseries},
  EdgeStyle/.append style = {->, bend left} }
  
  \usetikzlibrary{shapes.geometric}
\usetikzlibrary{positioning}

\title{Schr\"odinger's Ballot: Quantum Information and the Violation of Arrow's Impossibility Theorem}
\author{Xin Sun
\institute{Department of the Foundations of Computer Science, the John Paul II Catholic University of Lublin, Lublin, Poland}
\institute{Quantum Blockchains Inc., Pulawy, Poland}
\email{xin.sun.logic@gmail.com}
\and
Feifei He\thanks{Corresponding author}
\institute{Institute of Logic and Cognition, Sun Yat-sen University, Guangzhou, China}
\email{heff5@mail2.sysu.edu.cn}
\and
Piotr Kulicki
\institute{Department of the Foundations of Computer Science, the John Paul II Catholic University of Lublin, Lublin, Poland}
\institute{Quantum Blockchains Inc., Pulawy, Poland}
\email{kulicki@kul.pl}
\and
Mirek Sopek
\institute{MakoLab SA, Lodz, Poland}
\institute{Quantum Blockchains Inc., Pulawy, Poland}
\email{sopek@makolab.com}
}
\def\titlerunning{Quantum Information and the Violation of Impossibility Theorems of Voting in Social Choice Theory}
\def\authorrunning{Xin Sun, Feifei He, Piotr Kulicki  \&  Mirek Sopek}

\newtheorem{theorem}{Theorem}
\newtheorem{definition}{Definition}

\newtheorem{corollary}{Corollary}

\begin{document}
\maketitle

\begin{abstract}
In this paper we study Arrow's Impossibility Theorem in the quantum setting. Our work is based on the work of Bao and Halpern \cite{Bao17}, in which it is proved that the quantum analogue of Arrow's Impossibility Theorem is not valid. However, we feel unsatisfied about the proof presented in \cite{Bao17}. Moreover, the definition of Quantum Independence of Irrelevant Alternatives (QIIA) in \cite{Bao17} seems not appropriate to us. In this paper, we give a better definition of QIIA, which properly captures the idea of the independence of irrelevant alternatives, and a detailed proof of the violation of Arrow's Impossibility Theorem in the quantum setting with the modified definition. 
\end{abstract}

\section{Introduction}

Many voting protocols based on classical cryptography have been developed and successfully
applied since Chaum \textit{et al.} \cite{Chaum88}. However, the security of protocols based on classical cryptography is based on the unproven complexity of some computational algorithms, such as factoring of large numbers. The research in quantum computation shows that quantum computers are able to factor large numbers in a short time, which
means that classical protocols based on such algorithms are already insecure. To react to the risk posed by forthcoming quantum computers, a number of quantum voting protocols have been developed in the last decade \cite{Hillery06,Vaccaro07,Li08,Horoshko11,Li12,Jiang12,Tian16,Wang16,Thapliyal7,Sun19vote}.

While all these works have focused on the security problems of voting from a cryptographic perspective, Bao and Halpern \cite{Bao17} studied quantum voting from a social choice theoretic perspective by showing that the quantum analog of Arrow's Impossibility Theorem is violated in the quantum setting. 
Bao and Halpern's \cite{Bao17} idea and results are equally interesting.
However, we feel unsatisfied about the proof presented in  \cite{Bao17}. Moreover, the definition of the Quantum Independence of Irrelevant Alternatives (QIIA) in \cite{Bao17} seems not appropriate to us. 
In this paper, we give a better definition of QIIA, which properly captures the idea of independence of irrelevant alternatives, and a detailed proof of the violation of Arrow's Impossibility Theorem in the quantum setting with the modified definition. 

The structure of this paper is as follows. We review some background knowledge on classical and quantum voting in Section \ref{Background}.
In section \ref{Quantum Condorcet voting and Arrow's Impossibility Theorem} we introduce a voting rule called quantum Condorcet voting and proves that Arrow's Impossibility Theorem is violated by quantum Condorcet voting. We discuss related work in Section \ref{Related work} and conclude this paper with the plan for the future work in Section \ref{Conclusion and future work}. Some primitives of the quantum information theory which are used in this paper are collected in the appendix.

\section{Background}\label{Background}

\subsection{Classical voting system}
Now we succinctly review the theory of classical voting. A more detailed introduction to the classical voting and social choice theory can be found in Zwicker \cite{Zwicker16}, Pacuit \cite{Pacuit19} and Brandt \textit{et al.} \cite{Brandt16}.

Let $\mathcal{V}=\{v_1,\ldots,v_n\}$ be a finite set of (at least two)  voters and $\mathcal{C}=\{c_1,c_2,\ldots, c_n\}$ be a non-empty set of candidates. 
Each voter $v_i \in \mathcal{V}$ is endowed with preference $\succ_i$ over  $\mathcal{C}$. The preference $\succ_i$ is a binary relation on $\mathcal{C}$ that is  irreflexive, transitive, and complete. In other words, $\succ_i$ is a linear order on $\mathcal{C}$.
Let $\mathfrak{L}(\mathcal{C})$ denote the set of all linear orders on $\mathcal{C}$. A profile $\mathbf{R}=(R_1,\ldots,R_n) \in \mathfrak{L}(\mathcal{C})^{\mathcal{V} }$ is a vector of linear orders (i.e., preferences), where $R_i$ is the linear order supplied by voter $v_i$.
We write $\mathcal{V}^{\mathbf{R}}_{x\succ y }$ to denote the set of voters that rank candidate $x$ above candidate $y$ under profile $\mathbf{R}$.
A Social Welfare Function (SWF) is a function $F: \mathfrak{L}(\mathcal{C})^{\mathcal{V} } \mapsto \mathfrak{L}(\mathcal{C})$.
Two widely accepted properties of SWF are unanimity and independence of irrelevant alternatives.

\begin{definition}[Unanimity]
 A SWF $F$ satisfies the unanimity condition if, whenever all voters rank $x$ above $y$, then so does society: 
\begin{center}
$\mathcal{V}^{\mathbf{R}}_{x\succ y } = \mathcal{V}$ implies $(x,y)\in F(\mathbf{R})$.
\end{center}

\end{definition}

\begin{definition}[Independence of Irrelevant Alternatives (IIA)]
An SWF $F$ satisfies IIA if the relative social ranking of two candidates only depends on their relative voter rankings:

\begin{center}
$\mathcal{V}^{\mathbf{R}}_{x\succ y } = \mathcal{V}^{\mathbf{R}'}_{x\succ y }$ implies $(x,y)\in F(\mathbf{R})  \Leftrightarrow (x,y)\in F(\mathbf{R}')$.
\end{center}

\end{definition}

\noindent The intuition about IIA is that two ballot profiles that are \textit{similar} according to $(x,y)$ should produce the same ranking for $(x,y)$. This intuition will be used later to define the quantum analogue of IIA.

The celebrated Arrow's Impossibility Theorem states that any SWF that satisfies both unanimity and the IIA must also satisfy a property which any SWF should not satisfy: dictatorship.

\begin{definition}[Dictatorship]
An SWF $F$ satisfies dictatorship if there is a voter $v_i \in \mathcal{V}$ such that $F(\mathbf{R})= R_i$ for every profile $\mathbf{R}=(R_1,\ldots,R_n)$.

\end{definition}

\begin{theorem}[Arrow \cite{Arrow51}]
Any SWF for three or more candidates that satisfies unanimity and the IIA cannot satisfy nondictatorship.
\end{theorem}

\subsection{Quantum voting system}

Now we introduce our formalism of quantum voting system, which is similar to the formalism of Bao and Halpern \cite{Bao17}. 
In a quantum voting system with candidates $\mathcal{C}=\{c_1,c_2,\ldots, c_n\}$, we specify a Hilbert space $\mathsf{H}$ of which the dimension is $|\mathfrak{L}(\mathcal{C}) |$. That is, $\mathsf{H} = \mathbb{C}^{|\mathfrak{L}(\mathcal{C}) |}$. 
Every voter $v_i \in \mathcal{V}$ is associated with a Hilbert space $\mathsf{H}_i$ which is isomorphic to $\mathsf{H}$.  Every linear order $R \in \mathfrak{L}(\mathcal{C})$ is naturally viewed as basis vector $|R\rangle$ of $\mathsf{H}$. The basis $\mathsf{B} = \{ |R^1\rangle_i, \ldots, |R^{|\mathfrak{L}(\mathcal{C}) |}\rangle_i \}$ is called the preference basis for $\mathsf{H}_i$. 

Consider a pair  $(x,y)$ of candidates. $\mathsf{H}$ decomposes into subspaces associated with the possible relationships between $x$ and $y$. By $\mathsf{S}^{x \succ y }$, we denote the subspace spanned by the 
$\mathsf{B}$ elements that encode $x \succ y $ (e.g. $|x\succ y\succ z\rangle$, $|z\succ x\succ y\rangle$). We use $\Pi^{x \succ y}$ to denote the projector onto the subspace $\mathsf{S}^{x \succ y }$.

A quantum ballot of voter $v_i$ is a density operator $\rho \in D(\mathsf{H}_i)$. A quantum ballot profile is a density operator $\rho \in \mathsf{H}_1 \otimes \ldots \otimes \mathsf{H}_n$. A basis quantum ballot profile is a profile in which every component is a density operator of a basis vector.
A quantum social welfare function (QSWF) is linear map $\mathcal{E} :  D(\mathsf{H}_1 \otimes \ldots \otimes \mathsf{H}_n ) \mapsto D( \mathsf{H} ) $. The result of voting with quantum ballot profile $\rho$ is obtained by measuring $\mathcal{E}(\rho)$ on the preference basis. 

\begin{definition}[Quantum Unanimity \cite{Bao17}]

A QSWF $\mathcal{E}$ satisfies the sharp unanimity condition if it satisfies the following:

\begin{itemize}

\item For all quantum ballot profile $\rho$ and all pairs of candidates $(x,y)$, if $Tr( \Pi^{x \succ y} (Tr_{\neq i} (\rho)) ) =1$ for each voter $v_i$, then $Tr( \Pi^{x \succ y} (\mathcal{E} (\rho))) =1$.

\end{itemize}
\noindent
A QSWF $\mathcal{E}$ satisfies the unsharp unanimity condition if it satisfies the following:

\begin{itemize}
\item For all quantum ballot profile $\rho$ and all pairs of candidates $(x,y)$, if $Tr( \Pi^{x \succ y} (Tr_{\neq i} (\rho)) ) >0$ for each voter $v_i$, then $Tr( \Pi^{x \succ y} (\mathcal{E} (\rho))) >0$.

\end{itemize}
\noindent
A QSWF $\mathcal{E}$ satisfies the quantum unanimity condition if it satisfies both sharp and unsharp unanimity conditions.

\end{definition}

\begin{definition}[Quantum Independence of Irrelevant Alternatives (QIIA)]

A QSWF $\mathcal{E}$ satisfies the sharp IIA condition if it satisfies the following:
 
 \begin{itemize}

\item For all quantum ballot profile $\rho$ and $ \rho'$ and all pairs of candidates $(x,y)$, if $Tr( \Pi^{x \succ y} (Tr_{\neq i} (\rho)) ) = Tr( \Pi^{x \succ y} (Tr_{\neq i} (\rho')) )$ and $Tr( \Pi^{y \succ x} (Tr_{\neq i} (\rho)) ) = Tr( \Pi^{y \succ x} (Tr_{\neq i} (\rho')) )$ for each voter $v_i$, then 

$Tr( \Pi^{x \succ y} (\mathcal{E} (\rho))) =1  $ implies that $  Tr( \Pi^{x \succ y} (\mathcal{E} (\rho'))) =1$.

\end{itemize} 
 \noindent
 A QSWF $\mathcal{E}$ satisfies unsharp IIA if the following condition is satisfied:
 
\begin{itemize}

\item For all quantum ballot profile $\rho, \rho'$ and all pairs of candidates $(x,y)$, if $Tr( \Pi^{x \succ y} (Tr_{\neq i} (\rho)) ) = Tr( \Pi^{x \succ y} (Tr_{\neq i} (\rho')) )$ and $Tr( \Pi^{y \succ x} (Tr_{\neq i} (\rho)) ) = Tr( \Pi^{y \succ x} (Tr_{\neq i} (\rho')) )$
 for each voter $v_i$, then 
 
 $Tr( \Pi^{x \succ y} (\mathcal{E} (\rho))) >0  $ implies that $  Tr( \Pi^{x \succ y} (\mathcal{E} (\rho'))) >0$.

\end{itemize}
\noindent
A QSWF $\mathcal{E}$ satisfies the QIIA condition if it satisfies both sharp and unsharp  IIA conditions.

\end{definition}

\noindent
Note that QIIA in our definition is different from the QIIA in \cite{Bao17}. QIIA in \cite{Bao17} states that whether $\mathcal{E} (\rho)$ has support on $\mathsf{S}^{x \succ y }$ only depends on whether each $\rho_i$ has support on $\mathsf{S}^{x \succ y }$ and $\mathsf{S}^{y \succ x }$. More precisely, it states that for all quantum ballot profile $\rho, \rho'$ and all pairs of candidates $(x,y)$, if $Tr( \Pi^{x \succ y} (Tr_{\neq i} (\rho)) )>0$ iff  $  Tr( \Pi^{x \succ y} (Tr_{\neq i} (\rho')) )>0 $ and $Tr( \Pi^{y \succ x} (Tr_{\neq i} (\rho)) ) >0$ iff $ Tr( \Pi^{y \succ x} (Tr_{\neq i} (\rho')) ) >0$
 for each voter $v_i$, then  $Tr( \Pi^{x \succ y} (\mathcal{E} (\rho))) >0  $ implies that $  Tr( \Pi^{x \succ y} (\mathcal{E} (\rho'))) >0$.
 
 The intuition of QIIA is the same as the intuition of classical IIA: it states that two ballot profiles that are similar according to $(x,y)$ should produce the same ranking for $(x,y)$. It seems Bao and Halpern \cite{Bao17} considered two ballot profiles $\rho$ and $\rho'$ to be similar according to $(x,y)$ as long as $Tr( \Pi^{x \succ y} (Tr_{\neq i} (\rho)) )>0$ iff  $  Tr( \Pi^{x \succ y} (Tr_{\neq i} (\rho')) )>0 $ and $Tr( \Pi^{y \succ x} (Tr_{\neq i} (\rho)) ) >0$ iff $Tr( \Pi^{y \succ x} (Tr_{\neq i} (\rho')) ) >0$. To us this requirement is too weak. For example,  $\rho = 0.99 |  x \succ y \succ z\rangle\langle x \succ y \succ z| + 0.01 |y\succ  x \succ z\rangle  \langle y\succ  x \succ z |$ and $\rho' = 0.01 |  x \succ y \succ z\rangle\langle x \succ y \succ z| + 0.99 |y\succ  x \succ z\rangle  \langle y\succ  x \succ z |$ are similar according to $(x,y)$ in Bao and Halpern's definition, but intuitively they should not be. On the other hand, $\rho$ and $\rho'$ are not similar according to $(x,y)$ in our definition. Indeed, two profiles $\rho$ and $\rho'$ are similar according to $(x,y)$ in our definition only if  $Tr( \Pi^{x \succ y} (Tr_{\neq i} (\rho)) ) = Tr( \Pi^{x \succ y} (Tr_{\neq i} (\rho')) )$ and $Tr( \Pi^{y \succ x} (Tr_{\neq i} (\rho)) ) = Tr( \Pi^{y \succ x} (Tr_{\neq i} (\rho')) )$. We believe our definition of QIIA properly capture the idea of independence of irrelevant alternatives. That's why we use it to replace the QIIA of Bao and Halpern \cite{Bao17}.

\begin{definition}[Quantum Dictatorship \cite{Bao17}]

A QSWF $\mathcal{E}$ satisfies sharp dictatorship if there is a voter $v_i$ such that:
 \begin{itemize}

\item For all quantum ballot profile $\rho = ( \rho_1, \ldots, \rho_n)$ and all pairs of candidates $(x,y)$, $Tr( \Pi^{x \succ y} \rho_i) =1$  iff  $Tr( \Pi^{x \succ y} (\mathcal{E} (\rho))) =1$.

\end{itemize}
\noindent
 A QSWF $\mathcal{E}$ satisfies unsharp dictatorship if there is a voter $v_i$ such that:

\begin{itemize}

\item For all quantum ballot profile $\rho = ( \rho_1, \ldots, \rho_n)$ and all pairs of candidates $(x,y)$, $Tr( \Pi^{x \succ y} \rho_i)  >0$  iff  $Tr( \Pi^{x \succ y} (\mathcal{E} (\rho))) >0$.

\end{itemize}
\noindent
A QSWF $\mathcal{E}$ satisfies quantum dictatorship if it satisfies both sharp and unsharp  dictatorship.

\end{definition}

\section{Quantum Condorcet voting and Arrow's Impossibility Theorem}\label{Quantum Condorcet voting and Arrow's Impossibility Theorem}

We will use a special voting rule called Quantum Condorcet Voting $\mathcal{E}_{qcv}$ to disprove Arrow's Impossibility Theorem in the quantum setting. Since  $\mathcal{E}_{qcv} $ is a linear map from $D(\mathsf{H}_1 \otimes \ldots \otimes \mathsf{H}_n ) $ to $ D( \mathsf{H} ) $, we only need to specify how $\mathcal{E}_{qcv} $ operates on a basis quantum ballot profile.

\begin{definition}[Quantum Condorcet Voting]
Let $\rho_1 \otimes \ldots \otimes \rho_n$ be a basis quantum ballot profile. The Quantum Condorcet Voting  $\mathcal{E}_{qcv}$ operates in the following steps:

\begin{enumerate}

\item Calculates the Condorcet score of each candidate according to $\rho_1 \otimes \ldots \otimes \rho_n$. The Condorcet score of a candidate is the number of winning in pairwise comparison with other candidates. That is, for a candidate $x$, his Condorcet score $S_c(x)$ is $|\{y \in  \mathcal{C} : |\mathcal{V}^{\mathbf{R}}_{x\succ y }| \geq |\mathcal{V}^{\mathbf{R}}_{y\succ x }| \} |$ where $R$ is the classical ballot profile corresponding to $\rho_1 \otimes \ldots \otimes \rho_n$.

\item Generate a weak order $ \succeq$ over all candidates according to their Condorcet score. That is, $x\succeq y$ iff $S_c(x) \geq S_c(y)$.

\item Complete the weak order. That is, generate the set $\{\succ^1,\ldots, \succ^m \}$ in which each $\succ^i$ is a linear order that extends $\succeq$ and $\{\succ^1,\ldots, \succ^m \}$ contains all extensions of $ \succeq$. 

\item Transform the linear order into quantum states. That is, for $\{\succ^1,\ldots, \succ^m \}$ we create a quantum state $\sigma^1= \frac{1}{m} \sum\limits_{i} \sigma_i$, where each $\sigma_i$ is a basis ballot corresponds to $\succ^i$.

\item Give the minority a shot. For any candidate pair $(x,y)$ which is encoded by at least one $\rho_i$,
we spread an amount $\delta \in (0,1)$ of weight across
the $x\succ y$ subspace. That is, $\sigma^1$ is changed to $\sigma^2 = (1-k\delta)\sigma^1 + \delta \Pi^{x_1\succ y_1} +\ldots +  \delta \Pi^{x_k\succ y_k}$, where $(x_1,y_1),\ldots ,(x_k,y_k)$ ranges over all  candidate pairs that are encoded by at least one $\rho_i$. The parameter $\delta$ is required to satisfy that $\delta<\frac{1}{|\mathcal{C}|^2}$.

\item Enforce unanimity.  For any candidate pair $(x,y)$ which is encoded by all $\rho_i$, we project $\sigma^2$ onto the
$x\succ y$ subspace. That is, $\sigma^2$ is changed to $\sigma^3= \Pi^{x_k\succ y_k} \ldots \Pi^{x_1\succ y_1} \sigma^2 \Pi^{x_1\succ y_1} \ldots \Pi^{x_k\succ y_k} $, where $(x_1,y_1),\ldots ,(x_k,y_k)$ ranges over all  candidate pairs that are encoded by all $\rho_i$.

\end{enumerate}

\end{definition}

Both \textit{giving the minority a shot} and \textit{enforcing unanimity} are first introduced in Bao and Halpern \cite{Bao17}. While they may look strange at first sight, both of them will be useful in disproving Arrow's Impossibility Theorem.

\begin{theorem}
The Quantum Condorcet Voting $\mathcal{E}_{qcv}$  satisfies sharp unanimity.

\end{theorem}
\noindent 
\textit{Proof}: 
Let $\rho = \rho_1 \otimes \ldots \otimes \rho_n$ be a basis quantum ballot profile. If $Tr( \Pi^{x \succ y} (\rho_i) ) =1$ for each voter $v_i$, then each $\rho_i$ encodes $x \succ y$ since $\rho_i$ is a basis ballot. Then the projector $\Pi^{x\succ y}$ will be applied in the step of enforcing unanimity. Therefore, $Tr( \Pi^{x \succ y} (\mathcal{E}_{qcv} (\rho))) =1$.

Now let $\rho $ be a quantum ballot profile such that  $Tr( \Pi^{x \succ y} (Tr_{\neq i} (\rho)) ) =1$ for each voter $v_i$. Note that $Tr( \Pi^{x \succ y} (Tr_{\neq i} (\rho)) )=1 $ implies that $Tr_{\neq i} (\rho) = x^i_1|\phi^i_1\rangle \langle \phi^i_1| + \ldots + x^i_m|\phi^i_m\rangle \langle \phi^i_m|$ where each $\phi^i_j$ is a basis vector that encodes $x\succ y$ and $\sum\limits_{j} x^i_j=1$. Therefore, $\rho= x^1_1 \ldots x^n_1 |\phi^1_1 \otimes \ldots \otimes \phi^n_1 \rangle \langle   \phi^1_1 \otimes \ldots \otimes \phi^n_1 | +\ldots +  x^1_m \ldots x^n_m |\phi^1_m \otimes \ldots \otimes \phi^n_m \rangle  \langle  \phi^1_m \otimes \ldots \otimes \phi^n_m|$.
It then follows that $Tr( \Pi^{x \succ y} ( Tr_{\neq i}( |\phi^1_j \otimes \ldots \otimes \phi^n_j \rangle \langle |\phi^1_j \otimes \ldots \otimes \phi^n_j  |)) ) =1$ for all $i$. Then we know that $Tr( \Pi^{x \succ y} (\mathcal{E}_{qcv} (|\phi^1_j \otimes \ldots \otimes \phi^n_j \rangle \langle |\phi^1_j \otimes \ldots \otimes \phi^n_j  |))) =1$ because $|\phi^1_j \otimes \ldots \otimes \phi^n_j \rangle \langle |\phi^1_j \otimes \ldots \otimes \phi^n_j  |$ is a basis quantum ballot profile. Therefore, we have $Tr( \Pi^{x \succ y} (\mathcal{E}_{qcv} (\rho))) =  x^1_1 \ldots x^n_1  + \ldots +  x^1_m \ldots x^n_m = 1$.
\hfill $\square$

\begin{theorem}
The Quantum Condorcet Voting $\mathcal{E}_{qcv}$  satisfies unsharp unanimity.

\end{theorem}

\noindent 
\textit{Proof}: 
Let $\rho= \rho_1 \otimes \ldots \otimes \rho_n$ be a basis quantum ballot profile where each $\rho_i$ is a basis vector of $\mathsf{H}_i$. If $Tr( \Pi^{x \succ y} (\rho_i) ) >0$ for each voter $v_i$, then each $\rho_i$ encodes $x \succ y$ since $\rho_i$ is a basis ballot. Then the projector $\Pi^{x\succ y}$ will be applied in the step of enforcing unanimity. Therefore, $Tr( \Pi^{x \succ y} (\mathcal{E}_{qpv} (\rho))) =1 >0$.

Now let $\rho $ be a quantum ballot profile such that  $Tr( \Pi^{x \succ y} (Tr_{\neq i} (\rho)) ) >0$ for each voter $v_i$. Note that $Tr( \Pi^{x \succ y} (Tr_{\neq i} (\rho)) ) >0$ implies that $Tr_{\neq i} (\rho) = x_i|\phi_i\rangle \langle \phi_i| + \ldots$ for some basis vector $|\phi_i\rangle$ which encodes $x\succ y$ and $0 < x_i\leq 1$. Hence $\rho =  x_1 \ldots x_n |\phi_1 \otimes \ldots \otimes \phi_n \rangle \langle \phi_1 \otimes \ldots \otimes\phi_n| +\ldots  $. Note that $|\phi_1 \otimes \ldots \otimes \phi_n \rangle \langle \phi_1 \otimes \ldots \otimes\phi_n|  $ is a basis quantum ballot profile in which each $\phi_i$ encode  $x\succ y$. It then follows that 
 $Tr( \Pi^{x \succ y} (\mathcal{E}_{qcv} (|\phi_1 \otimes \ldots \otimes \phi_n \rangle \langle \phi_1 \otimes \ldots \otimes\phi_n| ))) =1$. From $0<  x_1 \ldots x_n  \leq 1$ we now know that   $Tr( \Pi^{x \succ y} (\mathcal{E}_{qpv} (\rho))) >0$.
\hfill $\square$

\begin{theorem}
The Quantum Condorcet Voting $\mathcal{E}_{qcv}$ satisfies sharp IIA.

\end{theorem}
\noindent 
\textit{Proof}: 
Let $\rho=\rho_1 \otimes \ldots \otimes \rho_n$ and $\rho'=\rho'_1 \otimes \ldots \otimes \rho'_n$ be two basis quantum ballot profiles. Assume $Tr( \Pi^{x \succ y} (\rho_i) ) = Tr( \Pi^{x \succ y}  (\rho'_i) )$ and $Tr( \Pi^{y \succ x} (\rho_i) ) = Tr( \Pi^{y \succ x}  (\rho_i') )$  for each voter $v_i$.
If $Tr( \Pi^{x \succ y} (\mathcal{E}_{qcv} (\rho))) =1  $, then we know $x \succ y$ is encoded by all $\rho_i$. For otherwise $\Pi^{y \succ x}$ will appear in $\sigma^2$ in the step of giving the minority a shot, making $Tr( \Pi^{x \succ y} (\mathcal{E}_{qcv} (\rho))) <1  $. 
Since $Tr( \Pi^{x \succ y} (\rho_i) ) = Tr( \Pi^{x \succ y}  (\rho'_i) )$ and $Tr( \Pi^{y \succ x} (\rho_i) ) = Tr( \Pi^{y \succ x}  (\rho_i') )$  for each voter $v_i$, we know that $x \succ y$ is encoded by all $\rho'_i$. Hence  $  Tr( \Pi^{x \succ y} (\mathcal{E}_{qcv} (\rho'))) =1$.

Now, let $\rho $ and $\rho' $ be quantum ballot profiles such that 
$Tr( \Pi^{x \succ y} (Tr_{\neq i} (\rho)) ) = Tr( \Pi^{x \succ y} (Tr_{\neq i} (\rho')) )$ and $Tr( \Pi^{y \succ x} (Tr_{\neq i} (\rho)) ) = Tr( \Pi^{y \succ x} (Tr_{\neq i} (\rho')) )$ for each voter $v_i$.
If $Tr( \Pi^{x \succ y} (\mathcal{E}_{qcv} (\rho))) =1  $, then we know $x \succ y$ is encoded by all $\rho_i$. For otherwise $\Pi^{y \succ x}$ will appear in $\sigma^2$ in the step of giving the minority a shot and 
$\Pi^{y \succ x}$ will not appear in $\sigma^3$ in the step of enforcing unanimity, making $Tr( \Pi^{x \succ y} (\mathcal{E}_{qcv} (\rho))) <1  $. Since $Tr( \Pi^{x \succ y} (\rho_i) ) = Tr( \Pi^{x \succ y}  (\rho'_i) )$ and $Tr( \Pi^{y \succ x} (\rho_i) ) = Tr( \Pi^{y \succ x}  (\rho_i') )$  for each voter $v_i$, we know that $x \succ y$ is encoded by all $\rho'_i$. Hence  $  Tr( \Pi^{x \succ y} (\mathcal{E}_{qcv} (\rho'))) =1$. 
\hfill $\square$

\begin{theorem}
The Quantum Condorcet Voting $\mathcal{E}_{qcv}$ satisfies unsharp IIA.

\end{theorem}
\noindent 
\textit{Proof}: 
Let $\rho=\rho_1 \otimes \ldots \otimes \rho_n$ and $\rho'=\rho'_1 \otimes \ldots \otimes \rho'_n$ be two basis quantum ballot profiles. Assume $Tr( \Pi^{x \succ y} (\rho_i) ) = Tr( \Pi^{x \succ y}  (\rho'_i) )$ and $Tr( \Pi^{y \succ x} (\rho_i) ) = Tr( \Pi^{y \succ x}  (\rho_i') )$  for each voter $v_i$.
 Now, if $Tr( \Pi^{x \succ y} (\mathcal{E} (\rho))) >0  $, then $y\succ x$ is not encoded by all candidates. Without loss of generality, let's assume $\rho_1$ encodes $x\succ y$ but not $y \succ x$. Then  $\rho'_1$ also encodes $x\succ y$ but not $y \succ x$. Hence  $  Tr( \Pi^{x \succ y} (\mathcal{E}_{qcv} (\rho'))) >0$.

Now, let $\rho $ and $\rho' $ be quantum ballot profiles such that 
$Tr( \Pi^{x \succ y} (Tr_{\neq i} (\rho)) ) = Tr( \Pi^{x \succ y} (Tr_{\neq i} (\rho')) )$ and $Tr( \Pi^{y \succ x} (Tr_{\neq i} (\rho)) ) = Tr( \Pi^{y \succ x} (Tr_{\neq i} (\rho')) )$ for each voter $v_i$.
If $Tr( \Pi^{x \succ y} (\mathcal{E}_{qcv} (\rho))) > 0  $, then we know $y \succ x$ is not encoded by all $\rho_i$. Since $Tr( \Pi^{x \succ y} (\rho_i) ) = Tr( \Pi^{x \succ y}  (\rho'_i) )$ and $Tr( \Pi^{y \succ x} (\rho_i) ) = Tr( \Pi^{y \succ x}  (\rho_i') )$  for each voter $v_i$, we know that $y \succ x$ is not encoded by all $\rho'_i$. Hence  $  Tr( \Pi^{x \succ y} (\mathcal{E}_{qcv} (\rho'))) >0$. 
 \hfill $\square$

\begin{theorem}
The Quantum  Condorcet Voting $\mathcal{E}_{qcv}$ does not satisfy sharp dictatorship.

\end{theorem}
\noindent 
\textit{Proof}: 
We will construct a ballot profile in which no candidate is a dictator.
Let $\{x,y,z\}$ be the set of candidates.
Let $\rho=\rho_1 \otimes \rho_2 \otimes \rho_3$ be a quantum ballot profile where  $\rho_1 = |  x \succ y \succ z\rangle \langle x \succ y \succ z |, \rho_2 = |    y \succ z \succ x \rangle  \langle  y \succ z \succ x | ,  \rho_3 = |z\succ  x \succ y\rangle  \langle z\succ  x \succ y | $.
Then the weak order generated by $\mathcal{E}_{qcv}$ according to the Condorcet score is $ x \equiv y \equiv z$. The completion of $ x \equiv y \equiv z$ is $\{ x \succ y \succ z,  x \succ z \succ y, y \succ x \succ z, y \succ z \succ x, z \succ x \succ y, z \succ y \succ x \}$. Therefore the quantum state generated in step 4 of quantum Condorcet voting is $\sigma^1= \frac{1}{6} ( | x \succ y \succ z  \rangle \langle x \succ y \succ z  | +  | x \succ z \succ y \rangle \langle   x \succ z \succ y | +   |  y \succ x \succ z  \rangle \langle   y \succ x \succ z |+   |  y \succ z \succ x \rangle \langle  y \succ z \succ x | + | z \succ x \succ y\rangle \langle  z \succ x \succ y| + | z \succ y \succ x\rangle \langle  z \succ y \succ x| )$. Therefore, we have $Tr( \Pi^{x \succ y} \rho_1) =1$ but $Tr( \Pi^{x \succ y} (\mathcal{E}_{qcv} (\rho))) <1$,  $Tr( \Pi^{y \succ z} \rho_2) =1$ but $Tr( \Pi^{y \succ z} (\mathcal{E}_{qcv} (\rho))) <1$, $Tr( \Pi^{z \succ x} \rho_3) =1$ but $Tr( \Pi^{z \succ x} (\mathcal{E}_{qcv} (\rho))) <1$. This violates sharp dictatorship. 
 \hfill $\square$

\begin{theorem}
The Quantum  Condorcet Voting $\mathcal{E}_{qcv}$ does not satisfy unsharp dictatorship.

\end{theorem}
\noindent 
\textit{Proof}: 
Consider again the profile constructed in the above proof. We have $Tr( \Pi^{z \succ x} (\mathcal{E}_{qcv} (\rho))) > 0$ but $Tr( \Pi^{z \succ x} \rho_1) \ngtr 0$, $Tr( \Pi^{x \succ y} (\mathcal{E}_{qcv} (\rho))) > 0$ but $Tr( \Pi^{x \succ y} \rho_2) \ngtr 0$, $Tr( \Pi^{y \succ z} (\mathcal{E}_{qcv} (\rho))) > 0$ but $Tr( \Pi^{y \succ z} \rho_3) \ngtr 0$. This violates unsharp dictatorship. 
 \hfill $\square$

\vspace{0.2cm}
By combining theorems 2-7 we conclude that Quantum Condorcet Voting satisfies Quantum Unanimity and the QIIA but does not Quantum Dictatorship. In other words, we can infer the following corollary:
\begin{corollary}
Arrow's impossibility theory is not valid in quantum voting.
\end{corollary}

\section{Related work}\label{Related work}

Most of the related work on quantum voting focus on the security of voting. 
The first quantum voting protocol was proposed by Hillery \textit{et al.} \cite{Hillery06}. They proposed two voting modes, namely traveling ballot and distributed ballot to ensure the security of voting. 
The protocol designed by Vaccaro \textit{et al.} \cite{Vaccaro07} uses 
entangled states to ensure that the votes are anonymous and to allow the votes to be tallied. The entanglement is distributed over separated sites; the physical inaccessibility of any site is sufficient to guarantee the anonymity of the votes.
Horoshko and Kilin \cite{Horoshko11} proposed a quantum anonymous
voting scheme based on Bell-state. Their protocol protects both the voters from a curious tallyman and all the participants from a dishonest voter in an unconditionally secure way.
Wang \textit{et al.} \cite{Wang16} proposed a quantum anonymous voting protocol assisted by two kinds of entangled quantum states. They provided a mechanism of opening and permuting the ordered votes of all the voters in an anonymous manner; any party who is interested in the voting results can obtain the voting result through a simple calculation. Their protocol possesses the properties of privacy, self-tallying, nonreusability, verifiability, and fairness at the same
time.

In our previous work \cite{Sun19vote} a simple voting protocol based on Quantum Blockchain was proposed. Despite its simplicity, our protocol satisfies the most important properties of secure voting protocols: is anonymous, binding, non-reusable, verifiable, eligible, fair and self-tallying. The protocol could also be implemented using presently available technology. One limitation of this protocol is that it works for only 2 candidates. In a recent paper \cite{Sun21vote} we overcame this limitation by realizing classical Condorcet voting on Quantum Blockchain.

\section{Conclusion and future work}\label{Conclusion and future work}

In this paper, we study Arrow's Impossibility Theorem in the quantum setting. We first modify the definition of QIIA such that it precisely captures the idea of independence of irrelevant alternatives. Then we present a detailed proof of the violation of Arrow's Impossibility Theorem with our modified definition.

In \cite{Sun21vote}, we have demonstrated that Condorcet voting on
Quantum Blockchain significantly simplifies the task of electronic voting and in the same time ensures many desired security properties.  In the future, we will further improve Quantum Condorcet Voting such that it has advantages for both security and the quality of the democratic processes. We will also investigate the validity of other theorems of classical social choice theory in the quantum setting. Those theorems include Sen's Theorem on the impossibility of a Paretian Liberal \cite{Sen70}, the Muller-Satterthwaite Theorem on surjective monotonicity \cite{Muller77} and the Gibbard-Satterthwaite Theorem on strategic manipulation \cite{Gibbard73}.
The third direction of research we are interested in is quantum logic for social choice. Modal logic has been used as a powerful tool to model and reason about social choice \cite{Troquard11,AgotnesHW11,CinaE16,Parmann21}. It is both natural and valuable to develop a quantum logic to model and reason about quantum social choice.

\newpage

\section*{Appendix: basics of quantum information}

Some primitives of quantum information which are used in this paper are collected in this appendix. The readers who are interested in quantum information are recommended to textbooks such as Yanofsky and Mannucci \cite{Yanofsky08}, Scherer \cite{Scherer19}, Nielsen and Chuang \cite{Nielsen10} and Watrous \cite{Watrous18}.

\begin{definition}[Hilbert space]

A (finite-dimensional) Hilbert space $\mathsf{H}$ is a 

\begin{enumerate}
\item complex vector space, that is,
\begin{center}
$\phi,\psi \in \mathsf{H}$ and $a,b \in \mathbb{C}$ $\Rightarrow$ $a \phi +b\psi \in \mathsf{H}$,
\end{center}
\item with a (positive-definite) scalar product $\langle \cdot | \cdot \rangle : \mathsf{H} \times \mathsf{H} \mapsto \mathbb{C}$ such that for all $\phi,\psi,\phi_1,\phi_2\in \mathsf{H}$ and $a,b \in \mathbb{C}$ 
\begin{enumerate}
\item $\langle \phi | \psi \rangle = \overline{\langle \psi | \phi \rangle}$
\item $\langle \phi | \phi \rangle \geq 0$

\item $\langle \phi | \phi \rangle = 0$ iff $\phi =0$

\item $\langle \psi | a \phi_1 +b \phi_2 \rangle = a\langle \psi |  \phi_1  \rangle + b\langle \psi |  \phi_2 \rangle$

\end{enumerate}
\end{enumerate}

\end{definition}

In quantum computation and quantum information, we only consider finite-dimensional Hilbert spaces. A vector $\phi$ of a Hilbert space is usually represented in the Dirac notion as $| \phi \rangle$ in quantum computing.
 A Hilbert space $\mathsf{H}$ induces a norm $\lVert.\rVert$ defined by $\lVert  \phi  \rVert = \sqrt{\langle \phi|\phi \rangle}$ for any $\phi \in \mathsf{H}$.

\begin{definition}[orthonormal basis and dimension]

An orthonormal basis $\{ |\phi_i\rangle \}$ for a Hilbert space $\mathsf{H}$ is a basis of $\mathsf{H}$ whose vectors are unit vectors and are orthogonal to each other, that is, for any $|\phi_i\rangle, |\phi_j\rangle$, $\lVert \phi_i \rVert = 1$ and $\langle \phi_i|\phi_j \rangle = 0$. The dimension of a Hilbert space is the number of vectors of an orthonormal basis.

\end{definition}

\begin{definition}[tensor product]

Given Hilbert spaces $V$ and $W$ of dimension $m$ and $n$ respectively, their tensor product, denoted $V \otimes W$, is a $mn$-dimensional space consisting of linear combinations of outer products $|v\rangle \otimes |w\rangle$ of vectors $|v\rangle = (v_1,v_2,\dots,v_m)^T \in V$ and $|w\rangle = (w_1,w_2,\dots,w_n)^T \in W$, where

\begin{equation}
  |v\rangle \otimes |w\rangle =
  \begin{bmatrix}
  v_1w_1  \\
  v_1w_2 \\
  \vdots \\
  v_m w_n
  \end{bmatrix}
\end{equation}

\end{definition}

\begin{definition}[subspace]

A subspace of a Hilbert space $V$ is a subset $W$ of $V$ such that $W$ is also a Hilbert space.

\end{definition}

\begin{definition}[operator]

A  linear map $A: \mathsf{H} \mapsto \mathsf{H}$ is called an operator on $\mathsf{H}$. 

\end{definition}
\noindent
We use $L(\mathsf{H})$ to denote the set of all operators on $\mathsf{H}$.

\begin{definition}[adjoint]

The operator $A^*: \mathsf{H} \mapsto \mathsf{H}$ that satisfies $\langle A^* \phi | \psi \rangle = \langle \phi |A \psi\rangle$ for all $\phi, \psi \in \mathsf{H}$ is called the adjoint operator to $A$.

\end{definition}

\begin{definition}[projector]

A projector of a Hilbert space $\mathsf{H}$ is a linear map $P: \mathsf{H} \mapsto \mathsf{H}$ such that $P^2 = P$ and $P^*=P$.

\end{definition}

Projectors are related to projective measurements in quantum mechanics. We use an operator $M$ to represent an observable of the quantum system being observed, with a decomposition $M = \sum_{m}P_{m}$, where $P_m$ is the projector onto the eigenspace of $M$ with eigenvalue $m$. The result of measuring the state $|\psi\rangle$ will be one of $M$'s eigenvalues, and the probability of getting result $m$ is $p(m) = \langle\psi|P_m|\psi\rangle$.

\begin{definition}[trace]

Let $\mathsf{H}$ be a Hilbert space and $\rho$ be an operator on $\mathsf{H}$. The trace of $\rho$ is defined by

$$ Tr(\rho) = \sum_{i} \langle i|\rho|i \rangle $$
\noindent
where $\{|i\rangle\}$ is an orthonormal basis of $\mathsf{H}$.

\end{definition}

\begin{definition}[positive semidefinite operator]

An operator $A: \mathsf{H} \mapsto \mathsf{H}$ is positive semidefinite
if it holds that $A = B^*B $ for some operator $B \in L(\mathsf{H})$.

\end{definition}

\begin{definition}[density operator]

A positive semidefinite operator $\rho$ on  $\mathsf{H}$ is a density operator if it holds that $\rho =\rho^*$ and $Tr(\rho) =1$.

\end{definition}


\begin{definition}[partial trace]
Suppose the composite system of two subsystems $A$ and $B$ is described by the density operator $\rho_{AB}$. The partial trace over $B$ is defined by

$$ \rho_A = Tr_B(\rho_{AB}) = \sum_{i}(I_A \otimes \langle i|)\rho_{AB}(I_A \otimes |i\rangle)$$
\noindent
where $\{|i\rangle\}$ is an orthonormal basis of the Hilbert space $\mathsf{H}_B$. $\rho_A$ is called the reduced density operator of the subsystem $A$. The partial trace over $A$ can be defined in a similar way.

\end{definition}

\newpage


\end{document}